\documentstyle[12pt]{article}
\newcommand{\D}{\displaystyle}
\newcommand{\varr}{{\stackrel{\rightarrow}{\varepsilon}}}

\newcommand{\h}{\hbar}
\newcommand{\pa}{\partial}
\newcommand{\var}{\varepsilon}

\begin{document}
\begin{center}
{\bf The Quantization of Free EM Fields:An Alternative Approach To Solve The Vacuum 
Catastrophe}\\
\vspace{0.5cm}

{\bf H. Razmi\\
and\\
A.H. Abbassi}

\vspace{0.5cm}

{\it \small Department of Physics, School of sciences \\
Tarbiat Modarres University, P.O.Box 14155-4838\\
Tehran, I.R. Iran }\\
\end{center}

\begin{abstract}
It is shown that the canonical quantum field theory of radiation based on
the field theoretical generalization of a recently proposed [1] commutation relation
between position and momentum operators of massless particles leads to zero
vacuum energy.This may be considered as a step toward the solution of the cosmological
constant problem at least for the electromagnetic (EM) fields contribution. \\
\end{abstract}

PACS number(s): 03.70.+k,11.10.Ef
\newpage
\subsection*{Introduction:}
In a recent work it has been discussed that the generally accepted 
commutation relation $[\hat{Q}_i,\hat{P}_j]=i\h \delta_{ij}\hat{1}$ is not a 
suitable commutation relation between position and momentum operators for
massless particles and the following relation has been proposed [1]:\\
\begin{equation}
[\hat{Q}_i,\hat{P}_j]=i\h c^2 H^{-2} \hat{P}_i \hat{P}_j \ \ \ \ \ (massless \  particles)
\end{equation}
It seems that the problem of photon position operator has been solved by means of 
this commutation relation and this can be considered as the first advantage
of it [2].\\
Here we are going to find the effect of the field theoretical generalization
of (1) in the canonical process of quantizing free EM fields (i.e. the canonical
quantum field theory of radiation).
\subsection*{The Canonical Quantum Field Theory Of Radiation:}
Now let us consider quantum field theory of radiation. For free 
electromagnetic fields we have;\\
$L=-\frac{1}{4} F_{\mu \nu} F^{\mu\nu}$\\
$F^{\mu \nu}=\partial^{\nu}A^{\mu}-\partial^{\mu}A^{\nu}$\\
$\Box A^{\mu}-\partial^{\mu} (\partial_{\nu} A^{\nu})=0$\\
where in Lorentz gauge(Fermi Lagrangian density); 
\begin{equation}
L_f=-\frac{1}{2} (\pa_{\nu}A_{\mu})(\pa^{\nu} A^{\mu})\hspace*{2cm}
\end{equation}
\hspace*{4cm}$\Box A^{\mu}=0$\\
and the momentum density is 
\begin{equation}
\pi_{\mu}=\frac{\pa L}{\pa \dot{A}^{\mu}}=-\frac{1}{c^2} \dot{A}_{\mu}
\end{equation}
The solution of the equation $\Box A^{\mu}=0$ can be expanded in a complete 
set of solutions of the wave equation. Fourier expansion by imposing the 
periodic boundary condition gives: 
\begin{equation}
A^{\mu} (x)=A^{\mu^+}(x) +  A^{\mu^-} (x)
\end{equation}
where
\begin{equation}
A^{\mu^+}(x)=\D{\sum_r}\D{\sum_{\vec k}} 
\left( \frac{\h c^2}{2V \omega} \right)^{\frac{1}{2}} \var^{\mu}_r (\vec k) 
\hat{a}_r (\vec k) e^{-ikx}
\end{equation}
\begin{equation}
A^{\mu^-}(x)=\D{\sum_r}\D{\sum_{\vec k}} 
\left( \frac{\h c^2}{2V \omega} \right)^{\frac{1}{2}} \var^{\mu}_r (\vec k) 
\hat{a}_r^+ (\vec k) e^{+ikx}
\end{equation}
where\\
$\omega=ck^0 =c|\vec k|$ and $A^{\mu}$ is appropriately normalized.\\
The summation over $r$, from $r=0$ to $r=3$, corresponds to the fact that for 
the field $A^{\mu}(x)$ there exists four linearly independent 
polarization states for each $\vec k$. These are described by the polarization 
vectors $\var_r^{\mu} (\vec k)$ which we choose to be real and satisfy the 
orthonormality and completeness relations:
\begin{equation}
\var_r(\vec k)\var_s(\vec k)=\var_{r \mu} (\vec k) \var_s^{\mu} (\vec k)=-\xi_r\delta_{rs}
\end{equation}
\begin{equation}
\D{\sum_r} \xi_r \var_r^{\mu} (\vec k) \var_r^{\nu} (\vec k) =-g^{\mu \nu}
\end{equation}
\begin{equation}
\left \{ \begin{array}{l}
\xi_0=-1,\xi_1=\xi_2=\xi_3=1\\
g_{00}=-g_{11}=-g_{22}=-g_{33}=1
\end{array} \right .
\end{equation}
A specific choice of polarization vectors in one given frame of reference 
often facilitates the interpretation. We shall choose these vectors as 
$$\var_0^{\mu}=n^{\mu}=(1,0,0,0) \ \ \ \ , \ \ \ \ \var_r^{\mu} (\vec k)=
(0, \varr_r (\vec k)), r=1,2,3$$
Where $\varr_1(\vec k)$ and $\varr_2 (\vec k)$ are usually orthogonal unit vectors
which are also orthogonal to $\vec k$, and $\varr_3 (\vec k)=\frac{\vec k}{|\vec k|}$. 
Since we are going to calculate the vacuum energy which is a Lorentz invariant 
and also gauge independent quantity, this special choice of polarization 
vectors doesn't restrict the validity of discussion. 

The Hamiltonian is: 
\begin{equation}
\hat{H}=\int d^3 x (\hat{\pi}_{\mu} \dot{A}^{\mu} -\hat{L})
\end{equation}
Using (3)-(6),(10) leads to: 
\begin{equation}
\hat{H}=\D{\sum_r} \D{\sum_{\vec k}} \left[ \frac{\h \omega}{2} \right] \xi_r 
\left [\hat{a}_r^+
 (\vec k) \hat{a}_r (\vec k)+\hat{a}_r (\vec k)\hat{a}_r^+(\vec k) \right]
\end{equation}
It is usual to impose the normal ordering condition and write $\hat{H}$ as 
\begin{equation}
\hat{H}=\D{\sum_r}\D{\sum_{\vec k}} (\h \omega) \xi_r [\hat{a}_r^+
(\vec k) \hat{a}_r (\vec k)]
\end{equation}
Normal ordering is only a way for removing the infinity of the vacuum energy 
and of course all excited states energies. In the cases that the absolute 
value of energy is not important the normal ordering will appear to be 
reasonable. But when the absolute value of energy is important, particularly 
in calculating the vacuum energy, the normal ordering does not seem to be 
reasonable. Here, we are going to show that by means of the field theoretical 
generalization of the new commutation relation (1), and without imposing the 
normal ordering condition, the undesired infinity in
 $\hat{H}$ will be disappeared. \\
Following the same procedure usually used to pass from quantum mechanics of 
discrete systems to canonical quantum field theory, the field theoretical 
generalization of (1) is found to be as 
\begin{equation}
\left[ \hat{A}^{\mu} (\vec x,t),\hat{\pi}_{\nu}(\vec x',t)\right]=i\h c^2 
\hat{H}^{-2} \hat{P}^{\mu} \hat{P}_{\nu} \delta(\vec x-\vec x')
\end{equation}
In which equal time commutation relation has been considered and in the right
hand side $\hat{H}, \hat{P}^{\mu}$ and $\hat{P}_{\nu}$ are Hamiltonian and 
components of four-momentum operators of the field. Here we should mention 
that after finding the proper commutation relations between $\hat{a}_r$ and 
$\hat{a}_r^+$'s, one can show that the relation (13) preserves the local 
property of the theory and the microcausality relation can be verified (see
section A). \\
Of course,if one worry about the dependence of (13) on the Hamiltonian and
linear momenta which are all involve integrals over all space,we can replace
$\hat{H}^{-2} \hat{P}^{\mu} \hat{P}_{\nu}$ by $\hat{T}^{00-2} \hat{T}^{0 \mu}
\hat{T}_0{\nu}$ where $\hat{T}^{\mu \nu}$ is the Maxwell energy-momentum
tensor which is a local quantity and is a function of space and time and all the
following calculations can be repeated by means of the above replacement without any different result.It is worthwile
to note that the equal time commutation relation is only a bridge for transiting
to find the proper commutators of $\hat{a}_r$ and $\hat{a}_r^+$'s.\\		                                                                     
Let us verify how the relation (13) affects the commutation relation between 
$\hat{a}_r$ and \\
$\hat{a}_r^+$. Using (3), (13) becomes: 
\begin{equation}
-\frac{1}{c^2} \left [\hat{A}^{\mu} (\vec x,t),\dot{\hat{A}}_{\nu}(\vec x',t) \right]=
i\h c^2 \hat{H}^{-2} \hat{P}^{\mu} \hat{P}_{\nu} \delta (\vec x-\vec x')
\end{equation}
or: 
$$-\frac{1}{c^2} \D{\sum_r} \D{\sum_{r'}} \D{\sum_{\vec k}} \D{\sum_{\vec k'}} 
\left[ \frac{\h c^2}{2V \omega} \right]^{\frac{1}{2}}
\left[ \frac{\h c^2}{2V \omega'} \right]^{\frac{1}{2}}
\var_r^{\mu} (\vec k)\var_{r' \nu} (\vec k') [\hat{a}_r(\vec k) e^{-ikx} +\hat{a}_r^+
(\vec k) e^{ikx}, $$
\begin{equation}
(-ik{'}^{0} 
 c)\hat{a}_{r'} (\vec k') e^{-ik' x'}+(ik{'}^{0} c)\hat{a}_{r'}^+ (\vec k') 
e^{ik'x'}]=i\h c^2 \hat{H}^{-2} \hat{P}^{\mu} \hat{P}_{\nu} \delta (\vec x-\vec x')
\end{equation}
Note that, $(t=t')$ and $(k^0 c=\omega)$. \\
The application of $\hat{P}^0 \hat{P}_0 =\frac{\hat{H}^2}{c^2}$ and 
$\D{\sum_{\vec k}} \frac{1}{\sqrt{V}} e^{i\vec k\cdot 
 \vec x} \frac{1}{\sqrt{V}} e^{-i\vec k \cdot 
\vec x'}=\delta (\vec x-\vec x')$ in the relation (15) for $\mu=v=0$, leads to the 
commutation relation 
\begin{equation}
\left [ \hat{a}_0(\vec k), \hat{a}_0^+ (\vec k')\right]=-\hat{1} \delta_{\vec k \vec k'}
\end{equation}
Furthermore, for the electromagnetic fields we have 
\begin{equation}
\D{\sum_{i=1}^3} \hat{P}^i \hat{P}_i=\frac{\hat{H}^2}{c^2}
\end{equation}
Therefore letting $\mu=v=i$ in (15) and summing over $i$ from $1$ to $3$, we 
find that the relation (15) cannot be valid unless we have:
\begin{equation}
\left[ \hat{a}_1(\vec k),\hat{a}_1^+ (\vec k') \right]+
\left[ \hat{a}_2(\vec k),\hat{a}_2^+ (\vec k') \right]+
\left[ \hat{a}_3(\vec k),\hat{a}_3^+ (\vec k') \right]= +\hat{1}\delta_{\vec k \vec k'}
\end{equation}
Of course, one may easily find that 
$\left[ \hat{a}_r(\vec k),\hat{a}_r (\vec k') \right]=
\left[ \hat{a}_r^+(\vec k),\hat{a}_r^+ (\vec k') \right]=0$ and all other 
commutators vanish for $r\neq r'$. \\
It is also worthwhile to mention that we can still introduce the Hilbert space 
of particle states (Fock space) consisting of vacuum and the related excited 
states and one should not worry about particle, here photon, interpretation of 
the theory(see section B). \\
For scalar photons, we have found the commutauion relation (16) which is the 
same as the result of the usual quantum field theory of radiation. A 
remarkable point is that the sign of the right hand side of (16) is negative 
and a comparison with the harmonic oscillator solution shows that, despite the 
usual interpretation, $\hat{a}_0(\vec k)$ and $\hat{a}_0^+ (\vec k)$ have the roles 
of creation and annihilation operators respectively. \\
Now insertion of the relations (16) and (18) in (11) leads to the following 
expression for the Hamiltonian
$$\hat{H}=\sum_{\vec k}(\h \omega) \left[ 
\hat{a}_1^+ (\vec k) \hat{a}_1 
(\vec k)+\hat{a}_2^+(\vec k)\hat{a}_2(\vec k)+\hat{a}_3^+(\vec k) \hat{a}_3(\vec k)- 
\hat{a}_0(\vec k)\hat{a}_0^+(\vec k)\right]$$
The related vacuum energy is then; 
$$\begin{array}{ll}
\langle 0 |\hat{H}| 0\rangle =\D{\sum_{\vec k}} (\h \omega) 
\left [ \left ( \langle 0 | \hat{a}_1^+ (\vec k) \hat{a}_1 (\vec k)+\hat{a}_2^+(\vec k) 
\hat{a}_2 (\vec k)+\hat{a}^+_3(\vec k)\hat{a}_3(\vec k)|0 \rangle 
-\langle 0|\hat{a}_0(\vec k) \hat{a}_0^+ (\vec k)|0 \rangle \right) \right]=0
\end{array}$$
because as stated before, for scalar photons $\hat{a}_0^+(\vec k)$ has the role 
of annihilation operator. \\
{\bf Conclusion:}\\
Without imposing the unreasonable constraint of normal ordering we have shown 
that the electromagnetic fields vacuum energy is zero and this can be 
considered as the solution of vacuum catastrophe [3]. It may be also 
considered as a step toward the solution of the cosmological constant problem 
[4].\\
There are many works on the solution of the cosmological constant problem 
e.g. supersymmetry [5], quantum cosmology [6-9], supergravity [10-12] 
and so on [13]. In all of these 
works there are either physically unknown assumptions and principles or 
mathematical difficulties such as unmeasurability of the path integral used in 
quantum cosmology. In this paper we have tried to solve the cosmological 
constant problem in a simple and physically reasonable way at least for the 
contribution of electromagnetic fields to the vacuum state. \\
However,two important questions remain to be answered. Is our result in 
challenge with the Casimir effect [14]? Does this approach destroy the experimentally
confirmed predictions of QED? \\
To answer the first question we should mention that in this study we have dealt with the free electromagnetic fields Lagrangian 
density,but in the case of the Casimir effect one should enter the effect of 
boundary conditions and probably the role of the other real and virtual 
particles. Also in that case the force between two conducting plates is to be 
calculated and therefore the relative value of energy is important,  
while in our calculations we are interested in the absolute value of vacuum 
energy.Our answer to the second question is "no" and the related discussion 
is presented in section C.\\ 
Finally we should mention that the cosmological constant may be 
considered as something different from quantum fields vacuum energy. For 
example some authors have tried to consider it as a constant of integration in 
a new theory of gravitation [15-16]. Anyway, we have shown that the 
electromagnetic fields vacuum energy vanishes and there is no infinity 
catastrophe in the state of vacuum corresponding to the electromagnetic 
fields.\\

\newpage
{\bf Section A:} \\
In order to verify the microcausality condition, we find the covariant 
commutator of the fields at two arbitrary points at first and then show that 
the result vanishes when the points have space-like separation. \\
For the points $x:(ct, \vec x)$ and $x':(ct',\vec x')$, the covariant commutation 
relation\\
between $\hat{A}^{\mu} (x)$ and $\hat{A}^{\nu} (x')$ is: 
$$\begin{array}{l} 
[\hat{A}^{\mu} (x), \hat{A}^{\nu} (x')]= 
\D{\sum_{r}} \D{\sum_{r'}} \D{\sum_{\vec k}}  \D{\sum_{\vec k'}} 
\left(\frac{ \h c^2}{2V \omega} \right)^{\frac{1}{2}}
\left(\frac{ \h c^2}{2V \omega'} \right)^{\frac{1}{2}}
\var^\mu_r(\vec k) \var_{r'}^{\nu}(\vec k') \\
\left [ \hat{a}_r(\vec k) e^{-ikx}+ \hat{a}_r^+ (\vec k) e^{ikx}, \hat{a}_{r'} 
(\vec k') e^{-ik' x'} +\hat{a}_{r'}^+(\vec k') e^{ik' x'} \right]
\end{array}$$
Using (16),(18) and taking the limit $V\rightarrow \infty$ for which we must 
substitue\\
$\D{\sum_{\vec k}} \frac{1}{V}$ by $\frac{1}{(2\pi)^3} \int d^3 k$, the above 
relation becomes:
$$[\hat{A}^{\mu} (x), \hat{A}^{\nu} (x') ]=\frac{\h c^2}{2(2\pi)^3} 
\sum_r \var^{\mu}_r \var^{\nu}_r [\hat{a}_r, \hat{a}_r^+] \int 
\frac{ e^{-ik(x-x')}-e^{ik(x-x')}}{\omega} d^3 k$$
But, the integral $\int \frac{e^{-ik (x-x')}-e^{ik (x-x')}}{\omega} d^3 k$ is 
the famous $\Delta$-function up to a multipilicative constant. We know that 
$\Delta(x-x')$ is a Lorentz invariant function and $\Delta(\vec x-\vec x',0)=0$. 
Therefore $\Delta(x-x')$ vanishes for $(x-x')^2<0$ and the final result is: \\
$[\hat{A}^{\mu} (x), \hat{A}^{\nu} (x')]=0$, for $(x-x')^2<0$\\
which is the microcausality condition. 

\newpage
{\bf Section B:}\\
Suppose we have the following relations:\\

$$\left[\hat{a}_1(\vec k),\hat{a}_1^+(\vec k')\right]=n_1\hat{1}\delta_{\vec k \vec k'}$$\\
$$\left[\hat{a}_2(\vec k),\hat{a}_2^+(\vec k')\right]=n_2\hat{1}\delta_{\vec k \vec k'}$$\\
$$\left[\hat{a}_3(\vec k),\hat{a}_3^+(\vec k')\right]=n_3\hat{1}\delta_{\vec k \vec k'}$$\\
The numbers $n_1$,$n_2$ and $n_3$ satisfy the following equation:\\
                          $$n_1+n_2+n_3=1$$\\
Now,we can introduce the following new operators:\\
     $$\hat{b}_1=\frac{\hat{a}_1}{\sqrt{n_1}}$$\\
         $$\hat{b}_2=\frac{\hat{a}_2}{\sqrt{n_2}}$$\\
         $$\hat{b}_3=\frac{\hat{a}_3}{\sqrt{n_3}}$$\\
It is obvious that by means of this new operators we can easily introduce the 
Hilbert space of particle states(Fock space) and the particle,photon,interpretation
is still there.\\
Of course,one may think that $n_1$, $n_2$ and $n_3$ can be functions of $\vec k$,...;
but if we go to one dimension we can check that this is impossible.Also,the 
independency of different polarizations and the space isotropicity of the problem
indicate that the numbers $n_1$,... cannot be functions of $\vec k$.

\newpage
{\bf Section C:}\\
In this section we want to explain that our approach doesn't destroy the
successful predictions of QED.The successful predictions of QED are based
on Feynman rules which are themselves based on the Feynman propagator (the
Feynman delta-function).
The Feynman propagator is the inverse of the operator "$\Box$" in phase
space and this operator is the operator of fields equation which are derived,
by means of the Lagrange equation,from the (Fermi) Lagrangian density.Here
we have worked with the usual Lagrangian density and so we expect our approach doesn't
destroy the Feynman rules.More clearly,our Feynman delta-function is essentially
equivalent to the usual QED delta-function (see section A).\\
In fact,QED is a perturbative theory of the quantum theory of free EM fields.
But the quantum theory of free EM fields has an infinity catastrophe in its
vacuum state.Here we have tried to remove this infinity without using the unreasonable
constraint of normal ordering and without damaging those successful parts of QED
which are experimentally confirmed.
  
\newpage 

\begin{center}
{\bf REFERENCES} 
\end{center}

\begin{itemize}
\item[1] H.Razmi,Report No. quant-ph/9811071 (1998).
\item[2] A.Shojai, M.Golshani, Annales de la Fondation Louis de Broglie,
{\bf 22}(4), 373 (1997).
\item[3] R.J. Adler, B. Casey and O.C. Jacob, Am. J. Phys. {\bf 63} (7), 680 
(1995). 
\item[4]  S.M. Carroll, W.H. Press and E.L. Turner, Annu. Rev. 
Astron. {\bf 30}, 499 (1992).  
\item[5] B. Zumino, Nucl. Phys. B {\bf 89}, 535 (1975). 
\item[6] S.W. Hawking, Phys. Lett. B {\bf 134}, 403 (1984). 
\item[7] S.W. Hawking, Nucl. Phys. B {\bf 239}, 257 (1984). 
\item[8] S. Coleman, Harvard University Preprint No. HUTP-88/A022.
\item[9] T. Banks, University of California, Santa Cruz, Preprint No. SCIPP
88/09.     
\item[10] E. Cremmer, B. Julia, J. Scherk, S. Ferrara, L. Girardello, and
P. van Nieuwenhuizen, Phys. Lett. B {\bf 79}, 231 (1978).
\item[11] E. Cremmer, B. Julia, J. Scherk, S. Ferrara, L. Girardello, and
P. van Nieuwenhuizen, Nucl. Phys. B {\bf 147}, 105 (1979).
\item[12] E. Witten, and J. Bagger, Phys. Lett. B {\bf 115}, 202 (1982).
\item[13] S. Weinberg, Rev. Mode. Phys. {\bf 61} (1), 1 (1989). 
\item[14] H.B.G. Casimir, Proc. K. Ned. Akad. Wet. {\bf 51}, 635 (1948). 
\item[15] J.J. Van der Bij, H. Van Dam, and Y. J. Ng, Physica {\bf 116A},
307 (1982).  
\item[16] W. Buchmuller, and N. Dragon, Phys. Lett. B {\bf 207}, 292 (1988). 
\end{itemize}

\end{document}